\documentclass[12pt,a4]{article}

\usepackage[T1]{fontenc}
\usepackage{graphicx}
\usepackage[dvips]{epsfig}
\usepackage{amssymb,amsfonts,amsmath}
\usepackage{tabularx}
\usepackage{enumerate}

\begin{document}

\centerline{\Large A Climate Network Based Stability Index for El Ni{\~n}o Variability}  
\bigskip
\centerline{ Qing Yi Feng and  Henk A. Dijkstra}
\bigskip
\centerline {\small Institute for Marine and Atmospheric research Utrecht (IMAU),}
\centerline {\small Department of Physics and Astronomy, Utrecht University, Utrecht, The Netherlands.}
\bigskip

{\bf 
Most of the existing prediction methods  gave a false alarm regarding the  
El Ni\~no event in 2014 \cite{CDB}.  A crucial aspect is currently limiting 
the success of such predictions \cite{Fedorov2003, Chen2004Predictability,  
Yeh2009}, i.e. the stability of the slowly  varying Pacific climate. This
property determines whether sea surface temperature perturbations will be amplified
by coupled ocean-atmosphere feedbacks or not.  The so-called Bjerknes stability 
index  \cite{Jin2006, Kim2011a, Kim2011b, Kim2014b} has been developed for this 
purpose, but its evaluation is severely constrained by data availability. Here we present 
a new promising background  stability  index based on complex network theory 
\cite{Tsonis:2006tk, Donges2009,  Feng2014b}.  This index  efficiently 
monitors the changes in spatial correlations  in the Pacific  climate  and 
can be evaluated by using only sea surface temperature data.  
}

The development of Pacific sea surface temperature (SST) conditions during  
the year 2014 has again  indicated  that  we lack a sufficiently detailed understanding 
of the El Ni\~no/Southern Oscillation (ENSO)  phenomenon  to  predict its developments 
up  to six  months ahead with adequate skill. The eastern Pacific equatorial heat 
content anomalies were very  large during March-May 2014 and models 
predicted strong El Ni\~no  conditions by the end of the year.  However, the 
atmospheric response  to the associated  SST
anomalies did not lead to a  strong feedback. As a consequence, 
equatorial  Pacific SST  anomalies remained relatively small in December
and decayed afterwards.  {The peak of the NINO3.4 index  (the area-averaged  SST  
anomalies  over  the region 120$^\circ$W-170$^\circ$W $\times$ 5$^\circ$S-
5$^\circ$N)}  in late November 2014 even did not  exceed 1.0$^\circ$C. Early 
in 2015, warming conditions appeared near the dateline leading to the current 
(weak) El Ni\~no conditions  \cite{CDB}. 

After more than thirty years of 
active research and an initial steep curve toward understanding the basic 
ENSO processes, the model prediction skill in 2014  was rather disappointing. 
Since ENSO has  large effects on the local climate in large parts of the world 
with severe impacts   on nature and society, it is important to 
(re)consider what processes may be missing or misrepresented in these 
models.    

The  Zebiak and Cane (ZC) model \cite{Zebiak1987} is thought to capture 
the basic processes of ENSO and is used in the suite of models  providing 
predictions of El Ni\~no variability \cite{CDB}.  Many  studies using this model  
have lead to  the delayed  oscillator view of ENSO,  where  positive Bjerknes 
feedbacks  are responsible for the amplification of SST  anomalies  and ocean  
adjustment provides a negative delayed feedback  \cite{Neelin1998, 
Jin1997b}.  The  strength of the feedbacks is represented by 
a coupling strength $\mu$, which is  proportional to the amount of  
wind-stress  anomaly per SST anomaly. 

In the ZC model (see Methods), the (steady or seasonal)  background climate  (provided by 
observations) becomes unstable to  when the strength of the coupled processes 
exceeds a critical value. The critical  boundary $\mu = \mu_c$ is,  in dynamical 
systems  theory,   referred to as a  Hopf bifurcation  (steady background state) or 
a  Neimark-Sacker bifurcation  (seasonal background state). When  
{$\mu > \mu_c$}, oscillatory motion 
develops spontaneously \cite{Fedorov2000, 
Bejarano2008} and the spatial pattern of the resulting variability is 
usually referred to as the ENSO mode. When  conditions are such that  
{$\mu < \mu_c$}, the   ENSO mode is damped  and can only be excited 
by noise  \cite{Penland1996, Burgers1999}.  These critical  boundaries $\mu_c$ 
have been explicitly  calculated for ZC-type  models  \cite{Jin1996a} 
and shown to involve the same oscillatory  ENSO mode for both cases.  

The prototype dynamical system displaying this behavior is the normal form of the 
stochastic Hopf bifurcation, which is discussed  in the section 1 of the Supplementary 
Information (SI). Supplementary Fig.~1 shows that the oscillatory behaviour can be 
excited by  noise even when the  background  steady state  is stable.  Such a  response 
has been found in typical ZC models   when stochastic wind forcing  is introduced 
\cite{Roulston2000a}. Hence, although  some  still consider  the noise driven and 
sustained ENSO variability   to be two different views, they are  actually easily 
reconciled: it just depends on  whether the background climate  is stable or unstable \cite{Dijkstra2013Book}. 

In global climate models (GCMs), both the background state 
and the  growth/decay of the ENSO mode are controlled by similar coupled 
processes  \cite{vanderVaart2000}. In addition,  both are also affected by processes 
outside of the  Pacific basin such as those at midlatitudes and in  the equatorial Indian 
and Atlantic  Oceans. In these models, the critical boundary 
separating stable  and unstable  regimes is not easy to identify. In addition, 
the model behavior is highly transient due to the  fluctuations at 
sub-annual  time scales (such as westerly wind bursts)  and 
external decadal time-scale processes such as a changing 
radiative forcing.  

One successful quantity to analyse the stability of the Pacific 
climate is the  Bjerknes stability (BJ) index  \cite{Kim2011a, Kim2011b, 
Kim2014b} which  is  based on the delayed  oscillator    framework 
\cite{Jin2006}.  However, the calculation of the 
BJ index requires a comprehensive dataset of the mean ocean currents,
the mean ocean upwelling{,} and the zonal and vertical gradients of the mean 
upper ocean temperature.  Moreover, determining the linear correlations 
between variables in the BJ index formulation requires relatively long time
series of data.  Although the BJ index can be applied to reanalysis 
data, it cannot be used  in cases when only SST observations  
(such as  in the period before the TAO/TRITON array) are available
or  when observational time series are relatively short. 

As the stability of the Pacific climate is key to improve the skill 
in future ENSO predictions,  a more practical index (than the BJ index) 
of the stability  of the background state  based on only SST data is urgently 
needed.  Here, we build on the success of  complex network approaches to 
efficiently monitor changes in spatial correlations of the atmospheric surface 
temperature \cite{Gozolchiani2011}, which  has been  exploited for ENSO 
prediction \cite{Ludescher2013}. Changes in spatial correlations 
due to stability changes of a background state (e.g. due to  critical slowdown) are 
also efficiently measured by topological changes in complex  networks  
\cite{Mheen2013,  Feng2014b}.  In approaching a critical  
boundary, spatial correlations of anomalies on the background state increase
leading to more coherence in the network. 

Here, we reconstruct interaction networks from SST data which are denoted by 
Pearson  Correlation Climate  Networks (PCCNs) \cite{Feng2014b}. Grid points in the 
longitude-latitude coordinate  system  over the equatorial Pacific form  the  
nodes of the network. Links between these nodes are  determined by zero-lag  
(significant) correlations, as measured through the Pearson  correlation (see 
Methods and section 2 of the SI) between  time series of SST anomalies  at the 
nodes.  As a measure of the coherence in the PCCN, we determine the number 
of links of each node, i.e. its degree. We propose that the  skewness  of the degree 
distribution, the degree skewness index $S_d$,  is an adequate measure of 
the stability of  the Pacific climate.  

To  validate the new index $S_d$, we  use the results from a fully-coupled
ZC model \cite{vanderVaart2000}.  In its standard setup,  the Hopf bifurcation 
occurs at $\mu_c = 3.0$ (Supplementary Fig.~2).   Time series of the SST field are 
generated using an additional red-noise wind forcing  \cite{Roulston2000a} 
(see Methods and section 3 in SI  for details on the model and the  red noise 
product) for different values of $\mu$ between $2.7$ and $3.4$. 
From each  540-month (45-year) simulations  the degree fields of the networks 
reconstructed are shown in Fig.~1 for four values of $\mu$.   When 
the coupling strength $\mu$ is increased  from the subcritical $\mu < \mu_c$ 
to the supercritical regime $\mu > \mu_c$,  more nodes in the region between 
220$^\circ$E  and 280$^\circ$E get a higher  degree. When the critical boundary 
is approached, the spatial pattern of the  anomalies is more and more controlled 
by the ENSO mode, leading to large-scale coherence which is  efficiently  measured 
by  the network degree field.  Another distinct change of the degree fields   (Fig.~1a-1d) 
is that the propagating  patterns of  equatorially symmetric  Rossby waves  are becoming 
more prominent  when the background state enters the supercritical regime. 

Histograms  of the degree fields (the degree distributions) for two different values of 
$\mu$  are plotted in Fig.~2a-2b.  For {$\mu = 2.7 < \mu_c$} (Fig.~{2}a), the  
degree distribution  is bimodal, where  the first peak represents the low degree  nodes  
in Fig.~1a   and the second peak (located near 250) {represents} the high degree 
nodes. When $\mu$ is 
increased (Fig.~2b) a peak at even higher degree occurs  because the ENSO 
mode becomes more dominant in the SST anomalies  when  the background 
climate  moves into  the  supercritical regime \cite{Mheen2013, Feng2014b}.  
The skewness of  the degree distribution $S_d$  is  monotonically  decreasing 
with  increasing $\mu$ ({blue curve} in Fig.~2c).  There is 
also a  good correlation between  $S_d$  and the variance of the {NINO3.4 index 
($Var_{NINO3.4}$, green curve)} as shown in  Fig.~2c. However, note that under high values 
of $S_d$ (stable climate, $\mu < \mu_c$), the amplitude of {NINO3.4} 
depends on  the noise (cf. Supplementary Fig. 1). 

For the ZC-model results here, the  BJ index (the black curve in Fig.~2d, 
details  of the calculation can be found in section 4 of the SI) is  
monotonously increasing with increasing $\mu$ and crosses the zero line 
at about $\mu = 3.1$ which is close to $\mu_c$.  This indicates that 
the BJ index reasonably (but not perfectly) monitors the stability of the background state 
for this case, which is in accordance with results in Jin \textit{et al.} \cite{Jin2006}. 
The values of different terms in the  BJ index (also shown in Fig.~2d) indicate that 
the thermocline feedback is  dominant  in destabilizing the background state. It 
has been shown in models \cite{Kim2011b, Kim2014b}  and reanalysis data 
\cite{Kim2014b} that this feedback has been dominant during past ENSO events.

Next, we reconstruct  PCCNs for the observed SST data (the HadISST dataset 
\cite{rayner2003}) and  focus on two major ENSO events over the  last 60 years, which 
peaked in 1982 and 1997. To show the different network properties for these individual 
ENSO events, we used monthly SST data of 4 years 
surrounding the peak of the event (two years before and one year after the year in 
which the peak occurred).  Figure~3a-3b indicate that the degree fields of the PCCNs 
reconstructed  for these two major ENSO events are substantially different. The pattern 
for the 1997 event displays a much larger area of high degree than that of  the 1982 event 
which is more localized and has a smaller degree. Comparing these to the patterns in  Fig. 1, 
one would interpret the (1982) 1997 event to be in the (sub) supercritical regime. 

To monitor  the changes in the stability of the Pacific climate,  we applied the stability 
index  $S_d$ to the HadISST dataset from December 1951 to November 2014 (63 years).  
We reconstructed a PCCN for  every 10 years of monthly SST data and computed its $S_d$.  
By implementing a sliding-window  strategy with a shift of one month, we obtained a time 
series of $S_d$  ({black} curve in Fig.~3c),  showing the variation of the stability over the 
Pacific background climate over the  last 60 years.   The {3-month} running mean 
of NINO3 index of the same period is plotted as the green curve in Fig.~3c. We also  marked 
the $S_d$ values  of  the 1982 and 1997 events, clearly   indicating  that the $S_d$ value 
at the onset  of the 1982 ENSO event (January-March  value of $S_d = 0.03$ in 1982) 
is much   higher than  the one for the 1997 event (January-March  value of $S_d = -0.37$ 
in  1997{)}.  Actually, the value of $S_d$ was overall  low during the early 90s with a global 
minimum  just before the 1997 event.  The high value of $S_d$ early 1982 indicates that noise 
must have had a large  influence on   the development  of the 1982 El Ni\~no event. The noise 
product (see Methods)  from the {Florida State University}  wind stress  indeed indicates 
that high noise variability was present  during this time  (See section 3 of the SI, in 
particular Supplementary Fig.~{3}). 
  
In summary, using techniques of complex  network theory we have developed a 
novel  index $S_d$ which  provides a simple and adequate 
measure of the  stability of the   Pacific  climate   both in models 
and observations.   From Fig.~3c, it is interesting to see that the $S_d$ index 
increased in  the beginning of the year 2014, indicating that the Pacific climate  
became more stable. This is consistent with the fact that although large subsurface 
temperature anomalies developed in March 2014,  there was no growth of SST  
anomalies because the thermocline feedback was  weak and no strong El Ni\~no 
event could develop.   In this way, our new index can provide an additional 
tool which can help improve El Ni\~no predictions in the future.  

\clearpage 
\section*{Methods}
\begin{itemize} 
\item[]  {\bf Model}: The model used for this study is 
a fully-coupled  variant of the original Zebiak-Cane (ZC) model \cite{vanderVaart2000}.  
The ocean component of this model is a reduced-gravity shallow-water model, 
and the atmospheric component is a linear, steady-state model \cite{Gill1980} 
forced by the sea-surface temperature (SST). In contrast to the original ZC model, 
this model is not an anomaly model around a prescribed climatological mean state, 
but the model itself generates the mean climate state and its variability.   More detailed 
information on this model can be found in  Dijkstra and Neelin \cite{Dijkstra1995p2} and
van der Vaart  \textit{et al.} \cite{vanderVaart2000}.  We use  model output of 11 
simulations at different values of the coupling strength $\mu$, (2.70, 2.80, 2.90, 
2.95, 2.98, 3.00, 3.02, 3.10, 3.15, 3.25, 3.40). 

\item[] {\bf Red Noise}: We followed the same procedure  of introducing the red noise 
into the wind-stress  forcing in the model   as  in Roulston and Neelin \cite{Roulston2000a}. 
We used the reconstructed data of Pacific SST for the period of 1978-2004 \cite{Smith1996}, 
and the Florida State University pseudo-wind-stress data for the same period  
\cite{Legler1988}. Details of the construction of the noise product and its effect 
on the behavior of the ENSO variability in the ZC model is given in section 3 
of the SI. 

\item[] {\bf Network Reconstruction}:  We reconstruct Climate Networks (CNs) for 
SST data sets  both from the model output and observations (HadISST dataset 
\cite{rayner2003}).  The model SST output is considered on a  domain of 
(140$^\circ$E, 280$^\circ$E) $\times$ (20$^\circ$S, 20$^\circ$N) on a 
30 $\times$ 31 grid, which provides a time series of 540 monthly SST anomaly  
fields (with respect  to the time mean).  The nodes of the each CN are the 
grid points where we have the SST data.  A `link'  between  two nodes is determined 
by  a significant correlation between their   SST anomaly time series measured 
here by the  Pearson correlation \cite{Tsonis:2006tk}.   For this study, each PCCN 
consists of  30 (longitude)$\times$31 (latitude)$ = 930$ nodes,  and  as a threshold 
for significance ($p < 0.05$) we choose the  threshold value $\tau=0.5$ .

\end{itemize} 

\section*{Correspondence}
Correspondence and requests for materials should be addressed to Q. Y. Feng
(email: Q.Feng@uu.nl). 

\section*{Author contribution}
Both authors designed the study, Q. Y. Feng performed most of the analysis, 
and both authors contributed to the writing of the paper. 

\section*{Acknowledgments}
We would like to acknowledge the support of the LINC project (no. 289447) 
funded by  the Marie-Curie ITN program (FP7-PEOPLE-2011-ITN) of EC. 
The authors thank Jonathan Donges, Norbert Marwan, Reik Donner (PIK, Potsdam),  
Avi Gozolchiani (Bar-Ilan University, Ramat-Gan), Hisham Ihshaish (VORtech, Delft), 
and Shicheng Wen (Tongji University, Shanghai) for technical support. 
QYF thanks Dewi Le Bars, Fiona R. van der Burgt, Lisa Hahn-Woernle, Alexis 
Tantet, and Claudia E. Wieners (IMAU, Utrecht) for constructive comments on the manuscript.

\clearpage 

\begin{figure}[t]
\begin{center}
	\includegraphics[width=1\textwidth]{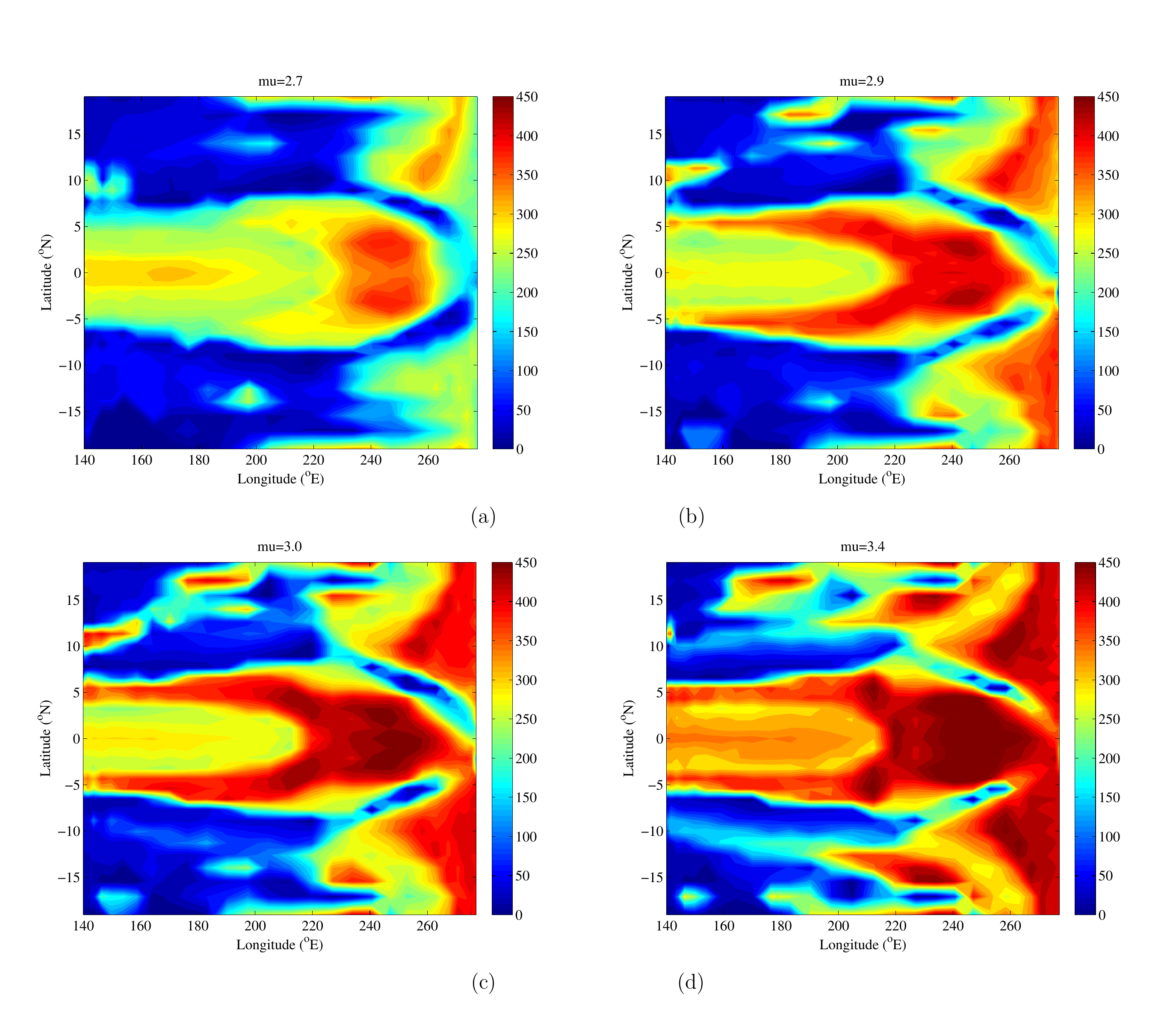} 
\end{center}
\caption{\it \small 
\textbf {(a)} Degree field of the Pearson Correlation Climate Network (PCCN) using a threshold 
$\tau = 0.5$ reconstructed from the modified ZC model data at the coupling strength $\mu=2.7$.   
\textbf {(b)} Same as \textbf {(a)} but at the coupling strength $\mu=2.9$. \textbf {(c)} Same as \textbf {(a)} 
but at the coupling strength $\mu=3.0$.  \textbf {(d)} Same as \textbf {(a)}  but at the coupling strength $\mu=3.4$.  }
\label{f:F2}
\end{figure}

\clearpage 
\begin{figure}[t]
\begin{center}
	\includegraphics[width=1\textwidth]{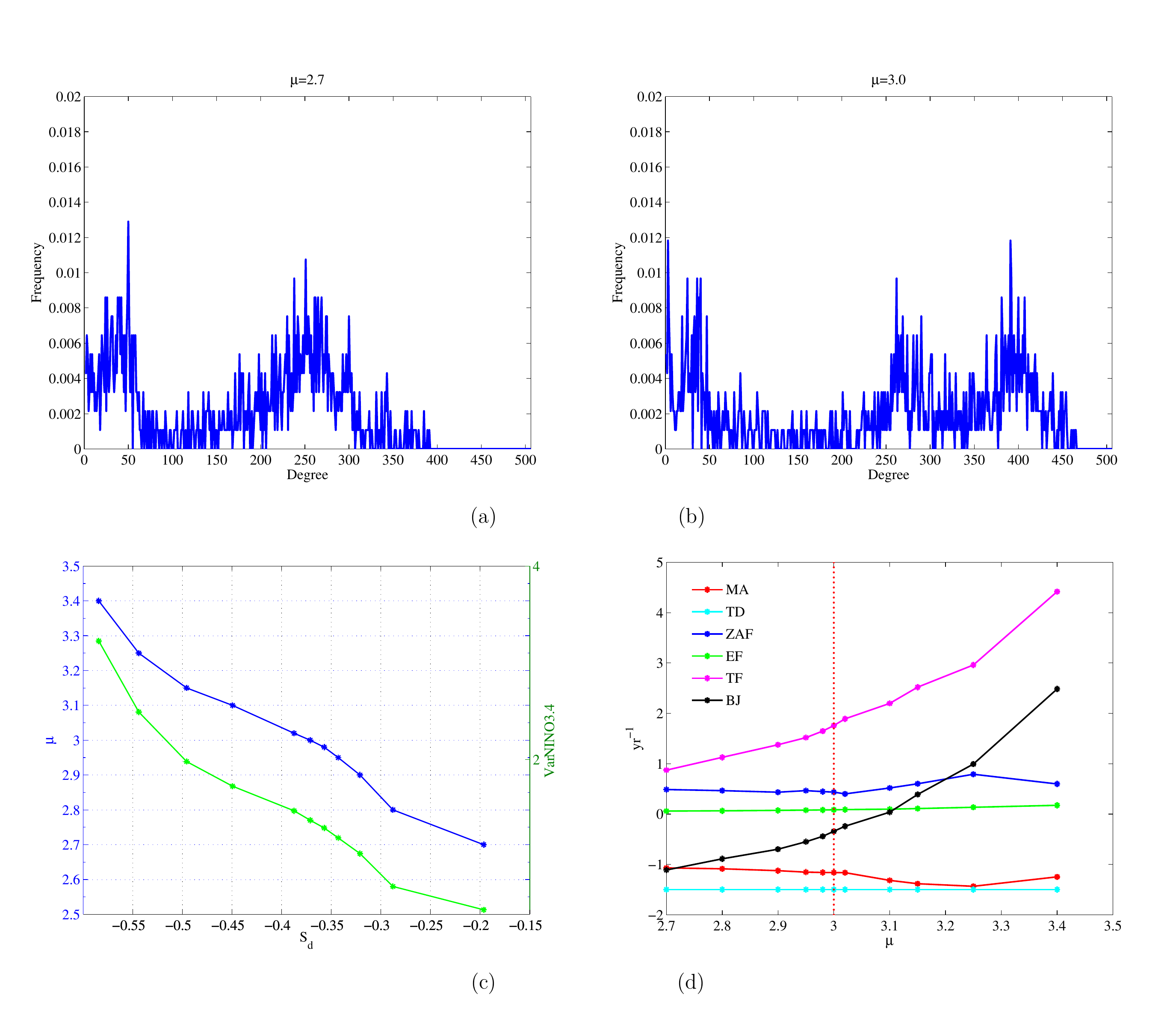} 
\end{center}
{\caption{\it \small 
\textbf {(a)} Degree distribution of the PCCN using a threshold 
$\tau = 0.5$ reconstructed from the modified ZC model data at the coupling strength $\mu=2.7$.   
\textbf {(b)} Same as \textbf {(a)} but at the coupling strength $\mu=3.0$. 
\textbf {(c)} The degree skewness index $S_d$ from the modified ZC model 
as a function of the coupling strength $\mu$ (blue) and the variance of NINO3.4 index $Var_{NINO3.4}$ (green). 
\textbf {(d)} The values of BJ index and its components 
from the modified ZC model at different coupling strength $\mu$ ($\mu_c=3.0$, red dash line):  
mean advection and upwelling (MA, red), thermal damping (TD, cyan), zonal advection feedback (ZAF, dark blue), Ekman 
pumping feedback (EF, green), thermocline feedback (TF, magenta), and total BJ index (BJ, black).
  }}
\label{f:F3}
\end{figure}

\clearpage 
\begin{figure}[t]
\begin{center}
	\includegraphics[width=1\textwidth]{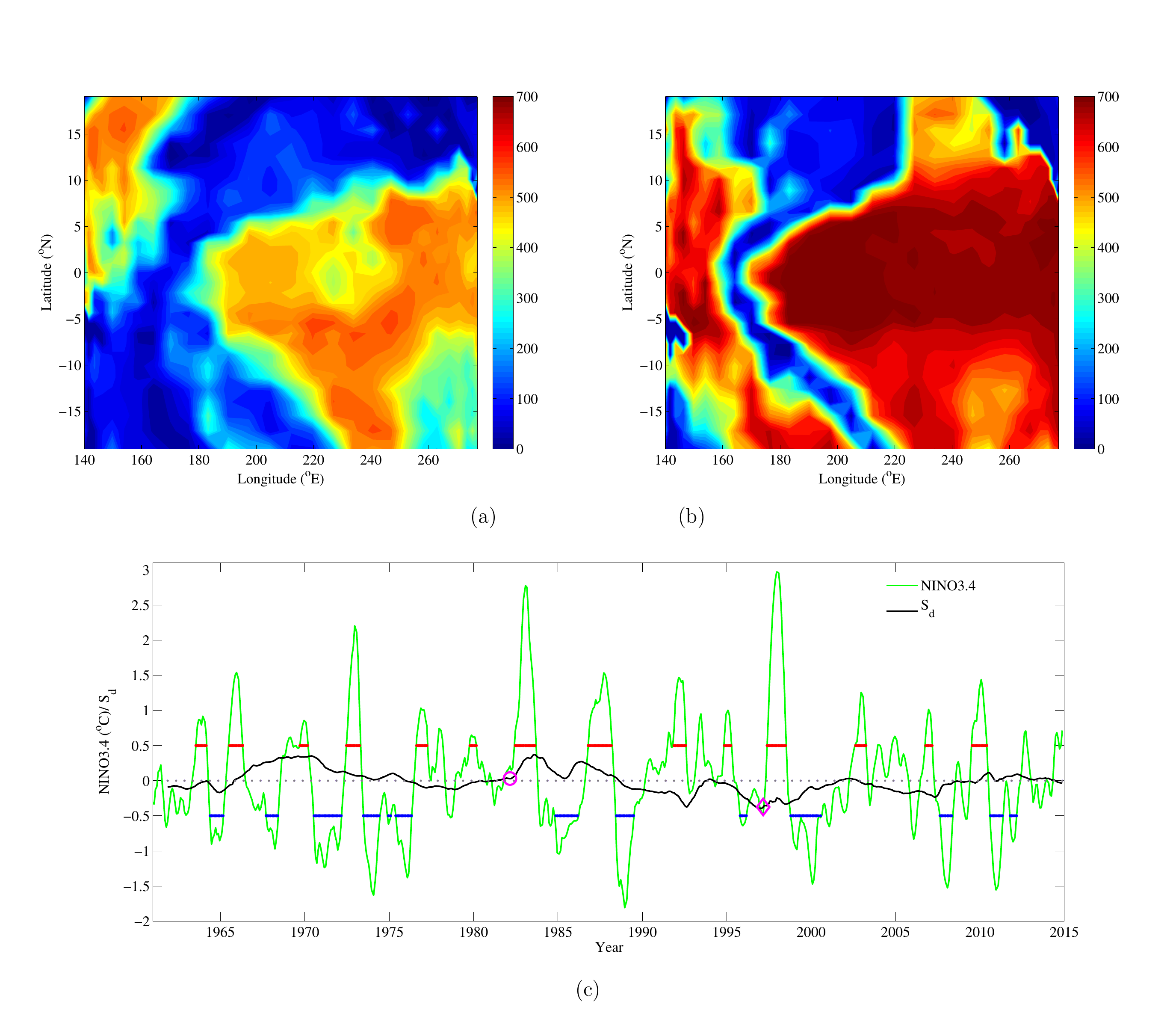} 
\end{center}
\caption{\it \small 
{\textbf {(a)} Degree field of the PCCN using a threshold 
$\tau = 0.5$ reconstructed from the observed SST of Jan 1980 to Dec 1983.   
\textbf {(b)} Same as \textbf {(a)} but of the observed SST of Jan 1995 to Dec 1998. 
\textbf {(c)} The  10-year sliding window degree skewness index $S_d$ (balck curve) 
and the 3-month running NINO3.4 index (green curve) 
from the observed SST of Dec 1951 to Nov 2014. 
The x-axis indicates the end time of the sliding window. 
The magenta circle indicates the mean $S_d$ value of the window ending 
at January-March 1982,  and the magenta diamond indicates that of January-March 
1997.  The red (blue) lines indicate El Ni\~no (La Ni\~na) events that 
a five consecutive 3-month running mean of NINO3.4 index 
that is above (below) the threshold of $+0.5^\circ$C ($-0.5^\circ$C). 
 }}
 
\label{f:F4}
\end{figure}

\end{document}